# Tri- & Tetra-Hyperbolic Iso-frequncy Topologies Complete Classification of Bi-Anisotropic Materials

**M. DURACH,**[1,*] **R. WILLIAMSON,**[1] **M. LABALLE**[1], **T. MULKEY**[1]

[1]*Department of Physics and Astronomy, Georgia Southern University, 65 Georgia Avenue, Statesboro, GA 30460*
*\*Corresponding author: mdurach@georgiasouthern.edu*

**Abstract.** We describe novel topological phases of iso-frequency k-space surfaces in bi-anisotropic optical materials – tri- and tetra-hyperbolic materials, which are induced by introduction of chirality. This completes the classification of iso-frequency topologies for bi-anisotropic materials, since as we show all optical materials belong to one of the following topological classes: tetra-, tri-, bi-, mono- or non-hyperbolic. We show that phase transitions between these classes occur in the k-space directions with zero group velocity at high k-vectors [Eq. (19)]. This classification is based on the sets of high-k polaritons (HKPs), supported by materials. We obtain the equation describing these sets [Eqs. (14), (17)] and characterize the longitudinal polarization impedance of HKPs [Eqs. (16), (18)]

Hyperbolic topologies spark imagination of the science fiction prosaists for almost a century [1]. In turn, the topology of iso-frequency k-surfaces in photonic materials fascinates the optics community workers. The known topologies include bounded k-surfaces such as spheres or ellipsoids, and unbounded k-surfaces – single- and double-leaf hyperboloids [2-4], and recently discovered bi-hyperboloids [5]. As can be seen from this list of k-surface topologies the main difference between them and the key to their classification are the high-k modes that propagate or not in these materials (see Fig. 1). The high-k modes are of primary interest in photonics and have already found applications for optical imaging with nanoscopic resolution using hyperlenses, emission control due to diverging optical density of high-k states and emission directivity control [6].

In this Letter we theoretically predict novel iso-frequency topology phases – tri- and tetra-hyperbolic materials and obtain an equation that describes the k-space directions in which the high-k modes exist in terms of the 36 material parameters of an arbitrary bi-anisotropic material [Eqs. (14), (17)]. We also employ a theorem due to Zeuthen (1873) [7], to show that our prediction of the tri- and tetra-hyperbolic topological phases completes the classification of bi-anisotropic materials, which means that an optical material belongs to one of the classes: tetra-, tri-, bi-, mono- or non-hyperbolic. The novel tetra- and tri-hyperbolic phases which we predict here are induced by introduction of chirality. Chirality induced modification of topology in the energy-momentum space was previously studied for mono-hyperbolic materials [8]; here we discuss the topology of iso-frequency surfaces in k-space.

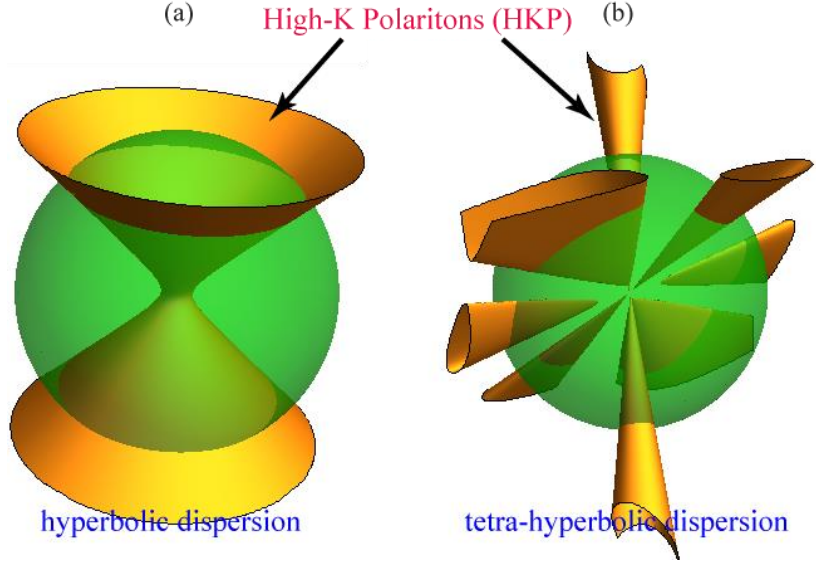

Fig. 1. Iso-frequency k-surface in k-space with High-k Polaritons (HKP) in a conventional mono-hyperbolic material (a) and a tetra-hyperbolic material (b) introduced in this paper and described in detail in the discussion of Figs. 3.

We show that in all high-k modes both electric and magnetic fields are polarized longitudinally and in this sense these modes are similar [9] to the bulk plasmon polaritons, bulk magnon polaritons, or longitudinal optical phonon polaritons, except that the high-k modes are short-wavelength. Therefore, below we call these modes as high-k polaritons (HKP).

For these HKPs we explicitly express the ratio between the longitudinal electric and magnetic field, i.e. longitudinal impedance $Z_l$, in terms of the 36 bi-anisotropic material parameters [Eqs. (16), (18)]. To accomplish this, we introduce the index of refraction operator $\widehat{N}$ [Eqs. (8)-(9)]. We obtain the direction-depended refraction indices explicitly in arbitrary bi-anisotropic material (Eqs. (10)-(12)).

Consider bi-anisotropic materials with 6x6 effective parameters matrix $\widehat{M} = \begin{bmatrix} \hat{\epsilon} & \widehat{X} \\ \widehat{Y} & \hat{\mu} \end{bmatrix}$, such that the constitutive relations are given by $\begin{pmatrix} \mathbf{D} \\ \mathbf{B} \end{pmatrix} = \widehat{M} \begin{pmatrix} \mathbf{E} \\ \mathbf{H} \end{pmatrix}$. It is known that the iso-frequency surface of the most generic bi-anisotropic material with arbitrary material parameters $\widehat{M}$ is a quartic surface in k-space given by [10]

$$f(\mathbf{k}, k_0) = \sum_{i+j+l+m=4}[\alpha_{ijlm} k_x^i k_y^j k_z^l k_0^m] = 0 \quad (1)$$

The topological asymptotic skeleton of the iso-frequency surfaces [Eq. (1)] can be found in the high-k limit $k \gg k_0$. The high-k states are tending to the conical k-surfaces given by

$$h(\mathbf{k}) = f(k \to \infty, k_0) = \sum_{i+j+l=4}[\alpha_{ijl0} k_x^i k_y^j k_z^l] = 0 \quad (2)$$

Below we establish the coefficients $\alpha_{ijl0}$ in terms of the material parameters $\hat{M}$ and the topological properties of the skeleton asymptotic surfaces of Eq. (2) [see Eq. (14)].

We start by considering Maxwell's equations for the amplitudes $\hat{\Gamma} = \begin{pmatrix} E \\ H \end{pmatrix}$ of the plane waves with wave vectors $k$ and frequencies $\omega = ck_0$ written as

$$k\hat{Q}\hat{\Gamma} = k_0 \hat{M}\hat{\Gamma}, \quad \hat{Q} = \begin{bmatrix} \hat{0} & \hat{R} \\ -\hat{R} & \hat{0} \end{bmatrix}, \quad \hat{R} = \hat{k} \times \tag{3}$$

Consider a wave propagating in the direction $\hat{k} = k/k = (q_x, q_y, q_z)$. Let us use the transformation $\hat{T}$:

$$\hat{T} = \begin{pmatrix} q_z q_x/q_\rho & q_z q_y/q_\rho & -q_\rho \\ -q_y/q_\rho & q_x/q_\rho & 0 \\ q_x & q_y & q_z \end{pmatrix} = \begin{pmatrix} \boldsymbol{q}_1 \\ \boldsymbol{q}_2 \\ \boldsymbol{q}_3 = \hat{\boldsymbol{k}} \end{pmatrix}$$

where $q_\rho = \sqrt{q_x^2 + q_y^2}$, and vectors $\boldsymbol{q}_i$ are the rows of matrix $\hat{T}$. Note that $\hat{T} = (\hat{T}^T)^{-1}$ and is a proper rotation $\det \hat{T} = 1$.

Let us apply transformation $\hat{T}$ to Maxwell's Eqs. (3)

$$k\tilde{Q}\tilde{\Gamma} = k_0 \tilde{M}\tilde{\Gamma}, \quad \tilde{\Gamma} = \hat{T}_6 \hat{\Gamma}, \quad \tilde{Q} = \hat{T}_6 \hat{Q} \hat{T}_6^{-1} = \begin{bmatrix} \hat{0} & \tilde{R} \\ -\tilde{R} & \hat{0} \end{bmatrix} \tag{4}$$

$$\tilde{R} = \begin{pmatrix} 0 & -1 & 0 \\ 1 & 0 & 0 \\ 0 & 0 & 0 \end{pmatrix}, \quad \tilde{M} = \hat{T}_6 \hat{M} \hat{T}_6^{-1} = \begin{bmatrix} \tilde{\epsilon} & \tilde{X} \\ \tilde{Y} & \tilde{\mu} \end{bmatrix}, \quad \hat{T}_6 = \begin{bmatrix} \hat{T} & 0 \\ 0 & \hat{T} \end{bmatrix} \tag{5}$$

According to Eq. (5) the transformed matrices $\tilde{a} = \tilde{\epsilon}, \tilde{\mu}, \tilde{X}, \tilde{Y}$ can be expressed in terms of the original $\hat{a} = \hat{\epsilon}, \hat{\mu}, \hat{X}, \hat{Y}$ as

$$\tilde{a}_{ij} = \boldsymbol{q}_i \cdot \hat{a} \cdot \boldsymbol{q}_j^T \tag{6}$$

The zeroes in the row at the bottom of matrix $\tilde{R}$ are due to the conditions $\boldsymbol{k} \cdot \boldsymbol{D} = 0$ and $\boldsymbol{k} \cdot \boldsymbol{B} = 0$. Thus, the longitudinal components of fields $E_z$ and $H_z$ can be decoupled from the transverse field components in the transformed coordinates as

$$\begin{pmatrix} E_z \\ H_z \end{pmatrix} = - \begin{pmatrix} \tilde{\epsilon}_{33} & \tilde{X}_{33} \\ \tilde{Y}_{33} & \tilde{\mu}_{33} \end{pmatrix}^{-1} \begin{pmatrix} \tilde{\epsilon}_{31} & \tilde{\epsilon}_{32} & \tilde{X}_{31} & \tilde{X}_{32} \\ \tilde{Y}_{31} & \tilde{Y}_{32} & \tilde{\mu}_{31} & \tilde{\mu}_{32} \end{pmatrix} \begin{pmatrix} E_x \\ E_y \\ H_x \\ H_y \end{pmatrix}, \tag{7}$$

while the transverse components satisfy the system of equations

$$k\begin{pmatrix} 0 & 0 & 0 & -1 \\ 0 & 0 & 1 & 0 \\ 0 & 1 & 0 & 0 \\ -1 & 0 & 0 & 0 \end{pmatrix}\begin{pmatrix} E_x \\ E_y \\ H_x \\ H_y \end{pmatrix} = k_0 \begin{pmatrix} \bar{\epsilon}_{11} & \bar{\epsilon}_{12} & \bar{X}_{11} & \bar{X}_{12} \\ \bar{\epsilon}_{21} & \bar{\epsilon}_{22} & \bar{X}_{21} & \bar{X}_{22} \\ \bar{Y}_{11} & \bar{Y}_{12} & \bar{\mu}_{11} & \bar{\mu}_{12} \\ \bar{Y}_{21} & \bar{Y}_{22} & \bar{\mu}_{21} & \bar{\mu}_{22} \end{pmatrix}\begin{pmatrix} E_x \\ E_y \\ H_x \\ H_y \end{pmatrix} = k_0 \bar{m} \begin{pmatrix} E_x \\ E_y \\ H_x \\ H_y \end{pmatrix} \quad (8)$$

$$\bar{m} = \begin{pmatrix} \tilde{\epsilon}_{11} & \tilde{\epsilon}_{12} & \tilde{X}_{11} & \tilde{X}_{12} \\ \tilde{\epsilon}_{21} & \tilde{\epsilon}_{22} & \tilde{X}_{21} & \tilde{X}_{22} \\ \tilde{Y}_{11} & \tilde{Y}_{12} & \tilde{\mu}_{11} & \tilde{\mu}_{12} \\ \tilde{Y}_{21} & \tilde{Y}_{22} & \tilde{\mu}_{21} & \tilde{\mu}_{22} \end{pmatrix} - \begin{pmatrix} \tilde{\epsilon}_{13} & \tilde{X}_{13} \\ \tilde{\epsilon}_{23} & \tilde{X}_{23} \\ \tilde{Y}_{13} & \tilde{\mu}_{13} \\ \tilde{Y}_{23} & \tilde{\mu}_{23} \end{pmatrix}\begin{pmatrix} \tilde{\epsilon}_{33} & \tilde{X}_{33} \\ \tilde{Y}_{33} & \tilde{\mu}_{33} \end{pmatrix}^{-1}\begin{pmatrix} \tilde{\epsilon}_{31} & \tilde{\epsilon}_{32} & \tilde{X}_{31} & \tilde{X}_{32} \\ \tilde{Y}_{31} & \tilde{Y}_{32} & \tilde{\mu}_{31} & \tilde{\mu}_{32} \end{pmatrix}$$

We rearrange this and introduce the index of refraction operator $\widehat{N}$:

$$\widehat{N}\begin{pmatrix} E_x \\ H_y \\ H_x \\ E_y \end{pmatrix} = n \begin{pmatrix} E_x \\ H_y \\ H_x \\ E_y \end{pmatrix}, \quad \widehat{N} = \begin{pmatrix} -\bar{Y}_{21} & -\bar{\mu}_{22} & -\bar{\mu}_{21} & -\bar{Y}_{22} \\ -\bar{\epsilon}_{11} & -\bar{X}_{12} & -\bar{X}_{11} & -\bar{\epsilon}_{12} \\ \bar{\epsilon}_{21} & \bar{X}_{22} & \bar{X}_{21} & \bar{\epsilon}_{22} \\ \bar{Y}_{11} & \bar{\mu}_{12} & \bar{\mu}_{11} & \bar{Y}_{12} \end{pmatrix} \quad (9)$$

The characteristic equation for the eigenvalue problem (9) is a quartic equation equivalent to Eq. (1)

$$n^4 - \mathrm{tr}(\widehat{N})\, n^3 - \xi n^2 - \zeta n + \det(\widehat{N}) = 0, \quad (10)$$

where

$$\xi = \frac{1}{2}\left(\mathrm{tr}(\widehat{N}^2) - \mathrm{tr}(\widehat{N})^2\right)$$

$$\zeta = \frac{1}{6}\left(2\,\mathrm{tr}(\widehat{N}^3) - 3\,\mathrm{tr}(\widehat{N}^2)\,\mathrm{tr}(\widehat{N}) + \mathrm{tr}(\widehat{N})^3\right)$$

Consider reciprocal materials. The material parameters are bound by $\hat{\epsilon}^T = \hat{\epsilon}$, $\hat{\mu}^T = \hat{\mu}$, $\hat{X}^T = -\hat{Y}$, which is true in the transformed coordinates for the elements of $\tilde{\epsilon}, \tilde{\mu}, \tilde{X}, \tilde{Y}$ and for the elements of $\bar{m}$ in Eqs. (8), (9)). In this case $\mathrm{tr}(\widehat{N}) = \mathrm{tr}(\widehat{N}^3) = 0$, which turns Eq. (10) into

$$n^4 - \tfrac{1}{2}\mathrm{tr}(\widehat{N}^2)\,n^2 + \det(\widehat{N}) = 0 \quad (11)$$

The roots of Eq. (11) are

$$n^2 = \tfrac{1}{4}\left\{\mathrm{tr}(\widehat{N}^2) \pm \sqrt{\left(\mathrm{tr}(\widehat{N}^2)\right)^2 - 16 \det \widehat{N}}\right\} \quad (12)$$

This is an explicit expression for the refraction indices of waves in arbitrary reciprocal materials, and this confirms our previous conclusion of Ref [11], that iso-frequency k-surface has reflection symmetry in reciprocal materials.

Let us turn to the asymptotic behavior at high k. If one of the eigenvalues of the index of refraction operator $\widehat{N}$ becomes infinite, then $\det \widehat{N} = n_1 n_2 n_3 n_4 \to \infty$, which is the product of all the eigenvalues $n_i$, also diverges. If the elements of the matrix $\widehat{M}$ are finite, then, according to the expression for matrix $\bar{m}$ [Eq. (8)], whose elements compose $\widehat{N}$ [Eq. (9)], such divergence is only possible for waves propagating in directions $\widehat{\boldsymbol{k}}$ such that

$$\Delta_\infty = \det\begin{pmatrix} \tilde{\epsilon}_{33} & \tilde{X}_{33} \\ \tilde{Y}_{33} & \tilde{\mu}_{33} \end{pmatrix} = 0. \tag{13}$$

Considering Eq. (6) and that $\boldsymbol{q}_3 = \widehat{\boldsymbol{k}}$, Eq. (13) can be rewritten as and equation of a quartic conical surface

$$h(\boldsymbol{k}) = 0, \tag{14}$$

where $h(\boldsymbol{k}) = (\boldsymbol{k}^T \hat{\epsilon} \boldsymbol{k})(\boldsymbol{k}^T \hat{\mu} \boldsymbol{k}) - (\boldsymbol{k}^T \hat{X} \boldsymbol{k})(\boldsymbol{k}^T \hat{Y} \boldsymbol{k})$.

Direct comparison shows that function $h(\boldsymbol{k})$ of Eq. (16) is identical to Eq. (2). Note that functions $a(\boldsymbol{k}) = (\boldsymbol{k}^T \hat{a} \boldsymbol{k}) = \sum_{ij} a_{ij} k_i k_j$, where $\hat{a} = \hat{\epsilon}, \hat{\mu}, \hat{X}, \hat{Y}$, are quadratic forms on k space.

If $\Delta_\infty \to 0$ then longitudinal components of fields $E_z$ and $H_z$ are much greater than transverse and $\boldsymbol{E} = E_z \widehat{\boldsymbol{k}}$, $\boldsymbol{H} = H_z \widehat{\boldsymbol{k}}$. Consequently Maxwell's equations can be written as

$$\boldsymbol{k} \cdot \boldsymbol{D} = (\boldsymbol{k}^T \hat{\epsilon} \boldsymbol{k}) E_z + (\boldsymbol{k}^T \hat{X} \boldsymbol{k}) H_z = 0, \tag{15a}$$

$$\boldsymbol{k} \cdot \boldsymbol{B} = (\boldsymbol{k}^T \hat{\mu} \boldsymbol{k}) H_z + (\boldsymbol{k}^T \hat{Y} \boldsymbol{k}) E_z = 0. \tag{15b}$$

Eqs. (15) have non-zero solution for the longitudinal fields $(E_z, H_z)$ only if Eq. (14) is met. Note that from $\boldsymbol{k} \times \boldsymbol{E} = k_0 \boldsymbol{B}$, $\boldsymbol{k} \times \boldsymbol{H} = -k_0 \boldsymbol{D}$ it follows that for HKPs $\boldsymbol{D} = \boldsymbol{B} = 0$, since HKP fields are purely longitudinal $\boldsymbol{E} = E_z \widehat{\boldsymbol{k}}$, $\boldsymbol{H} = H_z \widehat{\boldsymbol{k}}$.

An important characteristic of an HKP is its longitudinal impedance, which we introduce using Eqs. (15):

$$Z_l = \frac{E_z}{H_z} = -\frac{(\boldsymbol{k}^T \hat{X} \boldsymbol{k})}{(\boldsymbol{k}^T \hat{\epsilon} \boldsymbol{k})} = -\frac{(\boldsymbol{k}^T \hat{\mu} \boldsymbol{k})}{(\boldsymbol{k}^T \hat{Y} \boldsymbol{k})} \tag{16}$$

For reciprocal materials $(\boldsymbol{k}^T \hat{Y} \boldsymbol{k}) = -(\boldsymbol{k}^T \hat{X} \boldsymbol{k})$ and Eq. (14) breaks into two solution branches:

$$\sqrt{(\boldsymbol{k}^T \hat{\epsilon} \boldsymbol{k})(\boldsymbol{k}^T \hat{\mu} \boldsymbol{k})} \pm i(\boldsymbol{k}^T \hat{X} \boldsymbol{k}) = 0, \tag{17}$$

corresponding to impedances

$$Z_l = \pm i \sqrt{\frac{(k^T \hat{\mu} k)}{(k^T \hat{\epsilon} k)}} \tag{18}$$

The conventional condition of hyperbolicity requires the principle values of the dielectric permittivity tensor $\hat{\epsilon}$ to be of different signs [2-4]. Indeed, for such materials $h(k) = (k^T \hat{\epsilon} k)$ can only be zero if principle values of $\hat{\epsilon}$ have different signs. According to Eqs. (16) and (18) the longitudinal impedance for the HKP in these materials is $Z_l^{-1} = 0$. Similarly, for magnetic hyperbolic materials [12], the function from Eq. (16) is $h(k) = \epsilon \, (k^T \hat{\mu} k)$, assuming $\epsilon$ is a scalar, and the tensor $\hat{\mu}$ has to have principle values of different signs. The HKP waves have $Z_l = 0$ in magnetic hyperbolic materials [see Eqs. (16) and (18)].

It has been theoretically predicted in Ref. [7], that in absence of magnetoelectric coupling $\hat{X} = \hat{Y} = \hat{0}$ if both $\hat{\epsilon}$ and $\hat{\mu}$ have principal values of different signs at the same frequency $\omega$ then bi-hyperboloid iso-frequency k-surfaces are possible. As can be seen from Eq. (14), if $\hat{X} = \hat{Y} = \hat{0}$ then $h(k) = (k^T \hat{\epsilon} k)(k^T \hat{\mu} k)$ and, indeed, two hyperboloids in the k-surface can form corresponding to electric $Z_l^{-1} = 0$ branch with $(k^T \hat{\epsilon} k) = 0$ and magnetic branch $Z_l = 0$ with $(k^T \hat{\mu} k) = 0$. (cf. the discussion of Fig. 3)

Let us consider anisotropic magnetoelectric coupling and study the drastic changes to the topology of the iso-frequency surfaces it leads to. We first study materials, which are non-hyperbolic in absence of magnetoelectric coupling, i.e. the principal values of their $\hat{\epsilon}$ and $\hat{\mu}$ tensors have the same signs.

Consider a material with $\hat{\epsilon} = \epsilon \hat{1}$, $\hat{\mu} = \mu \hat{1}$ and $\hat{Y} = -\hat{X}^T = \text{diag}(i\kappa_1, i\kappa_1, i\kappa_2)$. Then Eq. (14) turns into $\left(q_\rho^2 \kappa_1 + q_z^2 \kappa_2\right)^2 = \epsilon\mu$, which shows that HKP propagate if $(\kappa_1^2 - \epsilon\mu)$ and $(\kappa_2^2 - \epsilon\mu)$ have opposite signs. Note that $\kappa = \kappa_1 = \kappa_2 = \pm\sqrt{\epsilon\mu}$ does not lead to formation of HKP, since in this case a 0/0 indeterminacy forms in Eq. (9), which resolves as 0, since for an isotropic chiral matrial $n = \sqrt{\epsilon\mu} \pm \kappa$ [13]. The formation of hyperbolic material here is due to anisotropic chirality tensor and is shown in Fig. 2(a).

If reciprocity is broken $\hat{X}^T \neq -\hat{Y}$, then $h(k) = \epsilon\mu - (k^T \hat{X} k)(k^T \hat{Y} k)$ and a bi-hyperboloidal k-surface may form. In Fig. 2(b) we plot k-surfaces for a material with $\hat{\epsilon} = \epsilon \hat{1}$, $\hat{\mu} = \mu \hat{1}$, $\hat{X} = -i \, \text{diag}(\kappa_1, \kappa_1, \kappa_2)$ and $\hat{Y} = i \, \text{diag}(\kappa_2, \kappa_2, \kappa_1)$ for the same values of $\kappa_1$ and $\kappa_2$ as in Fig. 2(a) and demonstrate the bi-hyperbolic dispersion.

In Fig. 2(c)-(d) we demonstrate that a material with anisotropic $\hat{\epsilon} = \text{diag}(1,1,5)$, $\hat{\mu} = \text{diag}(5,1,1)$, which is non-hyperbolic without magnetoelectric coupling, forms bi-hyperbolic and tri-hyperbolic phases if anisotropic chirality $\hat{Y} = -\hat{X}^T = \text{diag}(i\kappa_1, i\kappa_2, i\kappa_1)$ is added with $\kappa_1 = 2.9$ and $\kappa_2 = 0.9$ for Fig. 2(c) and $\kappa_1 = 2.3$ and $\kappa_2 = 1.1$ for Fig. 2(d).

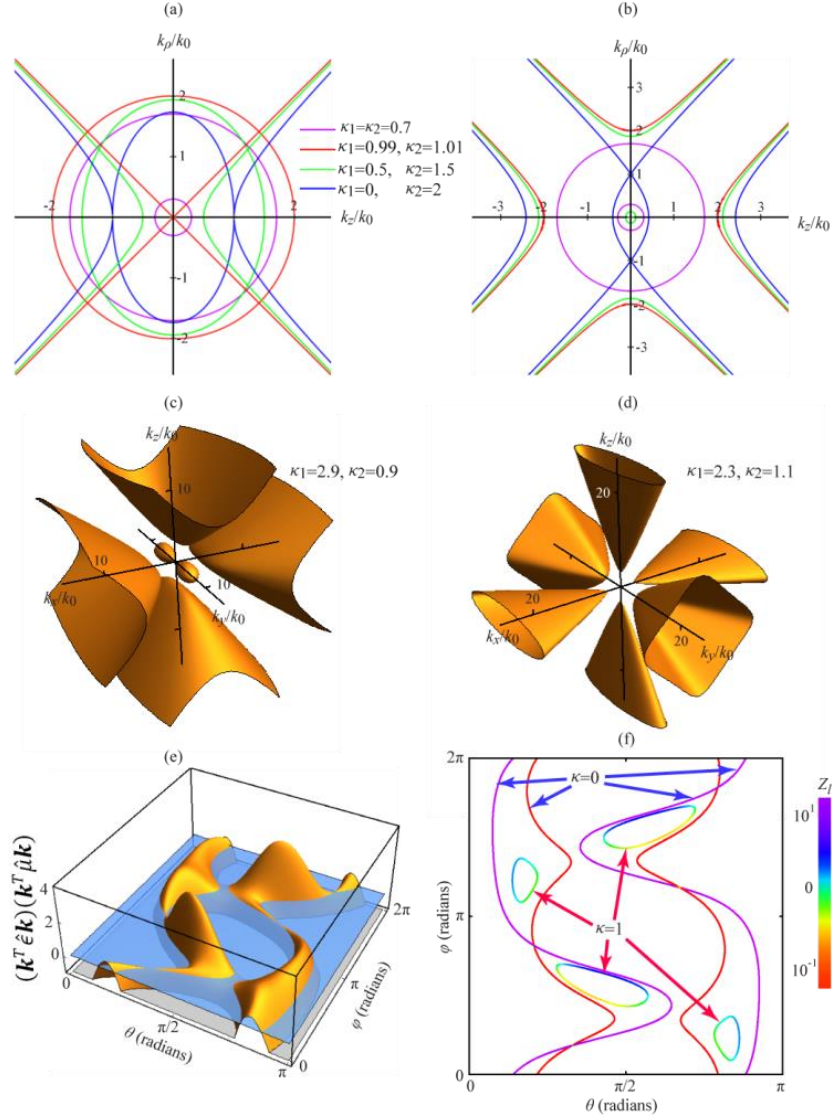

Fig. 2. (a-d) chirality inducing hyperbolicity in non-hyperbolic phases (material parameters detailed in text); (e-f) chirality-induced phase modifications in anisotropic materials. (e) Plot of the function $(\boldsymbol{k}^T \hat{\epsilon} \boldsymbol{k})(\boldsymbol{k}^T \hat{\mu} \boldsymbol{k})$ for the $\hat{\epsilon}, \hat{\mu}$ matrices used in Fig. 3(f) and Fig. 4.

Now let us turn to materials that are hyperbolic in absence of magnetoelectric coupling and consider how magnetoelectric coupling changes their topologies and HKPs. Below we consider a numerical example of matrix $M_1 = \begin{bmatrix} \hat{\epsilon} & 0 \\ 0 & \hat{\mu} \end{bmatrix}$ shown in the inset Fig. 3(d) (to the left of the black line). This matrix $M_1$ describes an anisotropic material with the dispersion transitional between hyperbolic and bi-hyperbolic. The HKP propagation directions for $M_1$ are illustrated in Fig. 2(e)-(f). In Fig. 2(e) we plot the function $(\boldsymbol{k}^T \hat{\epsilon} \boldsymbol{k})(\boldsymbol{k}^T \hat{\mu} \boldsymbol{k})$ (yellow), whose intersection with zero (blue) corresponds to the HSK propagation directions. These directions are also plotted in Fig. 2(f) for to electric $Z_l^{-1} = 0$ branch with $(\boldsymbol{k}^T \hat{\epsilon} \boldsymbol{k}) = 0$ (purple) and magnetic branch $Z_l = 0$ with

$(\mathbf{k}^T\hat{\mu}\mathbf{k}) = 0$ (red), which corresponds to the color scale for $Z_l$ shown to the right. Let us turn to a material characterized by matrix $\widehat{M} = \widehat{M}_1 + i\kappa\widehat{M}_2$, where $M_2 = \begin{bmatrix} \hat{0} & \hat{1} \\ -\hat{1} & \hat{0} \end{bmatrix}$ is responsible for magnetoelectric coupling. The change in the HKP propagation directions with increasing $\kappa$ from zero corresponds to raising the level of the blue plane in Fig. 2(e) to level $\kappa^2$, which confines the HKP states to the region with positive $(\mathbf{k}^T\hat{\epsilon}\mathbf{k})(\mathbf{k}^T\hat{\mu}\mathbf{k})$. The HKP propagation directions for $\kappa = 1$ are shown in Fig. 2(f) with 4 disconnected curves, which corresponds to bi-hyperbolic dispersion. These connectivity curves are color-coded corresponding to $Z_l$ of the HKP states.

Let us consider material described by matrix $\widehat{M} = \widehat{M}_1 + i\kappa\widehat{M}_2$ with the individual matrices $M_1, M_2$ plotted on Fig. 3(d). Changing parameter $\kappa$ leads to topological transitions between bi-hyperbolic [Fig. 3(a),(e)], tri-hyperbolic [Fig. 3(b),(g)] and tetra-hyperbolic [Fig. 3(c),(i)] phases. The progression of the phase changes is demonstrated in Figs. 3 (e)-(j). In these panels the HKP states for $\kappa = 0$ satisfying $(\mathbf{k}^T\hat{\epsilon}\mathbf{k})(\mathbf{k}^T\hat{\mu}\mathbf{k}) = 0$ are outlined in black, which is the same curves as in Fig. 2(f). The directions for which $\kappa = 0$ and $(\mathbf{k}^T\widehat{X}\mathbf{k}) = 0$ [brown] satisfy Eqs. (14) and (17) for all $\kappa$ and serve as the framework for the topological transitions. These intersection points in Figs. 3(e)-(j) have invariant impedances $Z_l^{-1} = 0$ for electric branch with $(\mathbf{k}^T\hat{\epsilon}\mathbf{k}) = 0$ and $Z_l = 0$ for magnetic branch.

The HKP states for the $\kappa$ noted in panels (e)-(j) are color coded in accordance with their $Z_l$ as indicated to the right of Fig. 3(f). The HKP for non-zero $\kappa$ are positioned in the $(\mathbf{k}^T\hat{\epsilon}\mathbf{k})(\mathbf{k}^T\hat{\mu}\mathbf{k}) > 0$ regions between the $\kappa = 0$ lines, in accordance with Eqs. (14) and (17), and touch these lines at the intersections of $\kappa = 0$ and $(\mathbf{k}^T\widehat{X}\mathbf{k}) = 0$. The HKP states group into several disconnected curves, the number of which characterizes the topology of the phase. Note that each curve is split into two parts by the $(\mathbf{k}^T\widehat{X}\mathbf{k}) = 0$ curves at their intersections with $\kappa = 0$. The resulting two subcurves of HKP states correspond to two different branches of Eq. (17).

Addition of small $\kappa$, leads to transition to the bi-hyperbolic phase (4 disconnected curves) as shown in Fig. 3(e). At $\kappa = 0.3$ [Fig. 3(f)] a phase transition occurs from the bi-hyperbolic to the tri-hyperbolic phase (6 curves) in the k-space direction marked by the red arrows. In this direction the group velocity, which is normal to the k-surface becomes zero

$$\mathbf{v}_g = \nabla_\mathbf{k}\omega \propto \nabla_\mathbf{k} f(\mathbf{k}, k_0) = \nabla_\mathbf{k} h(\mathbf{k}) = 0 \qquad (19)$$

In general, we identify the condition (19) with topological phase transitions. The transition from the tri-hyperbolic [Fig. 3(g)] to the tetra-hyperbolic phase [8 curves in Fig. 3(i)] occurs at $\kappa = 0.666$ [Fig. 3(h)]. The k-space direction in which the transition occurs satisfies Eq. (19) and is marked by the red arrows. For large $\kappa$ the HKP states follow closely the $(\mathbf{k}^T\widehat{X}\mathbf{k}) = 0$ curves between their intersections with $\kappa = 0$. Therefore, in the general case of arbitrary matrices $\widehat{M}_1$ and $\widehat{M}_2$ the number of disconnected sections of $(\mathbf{k}^T\widehat{X}\mathbf{k}) = 0$ between their intersections with $\kappa = 0$ can serve as the prediction of the topological phase of the iso-frequency surfaces for large $\kappa$.

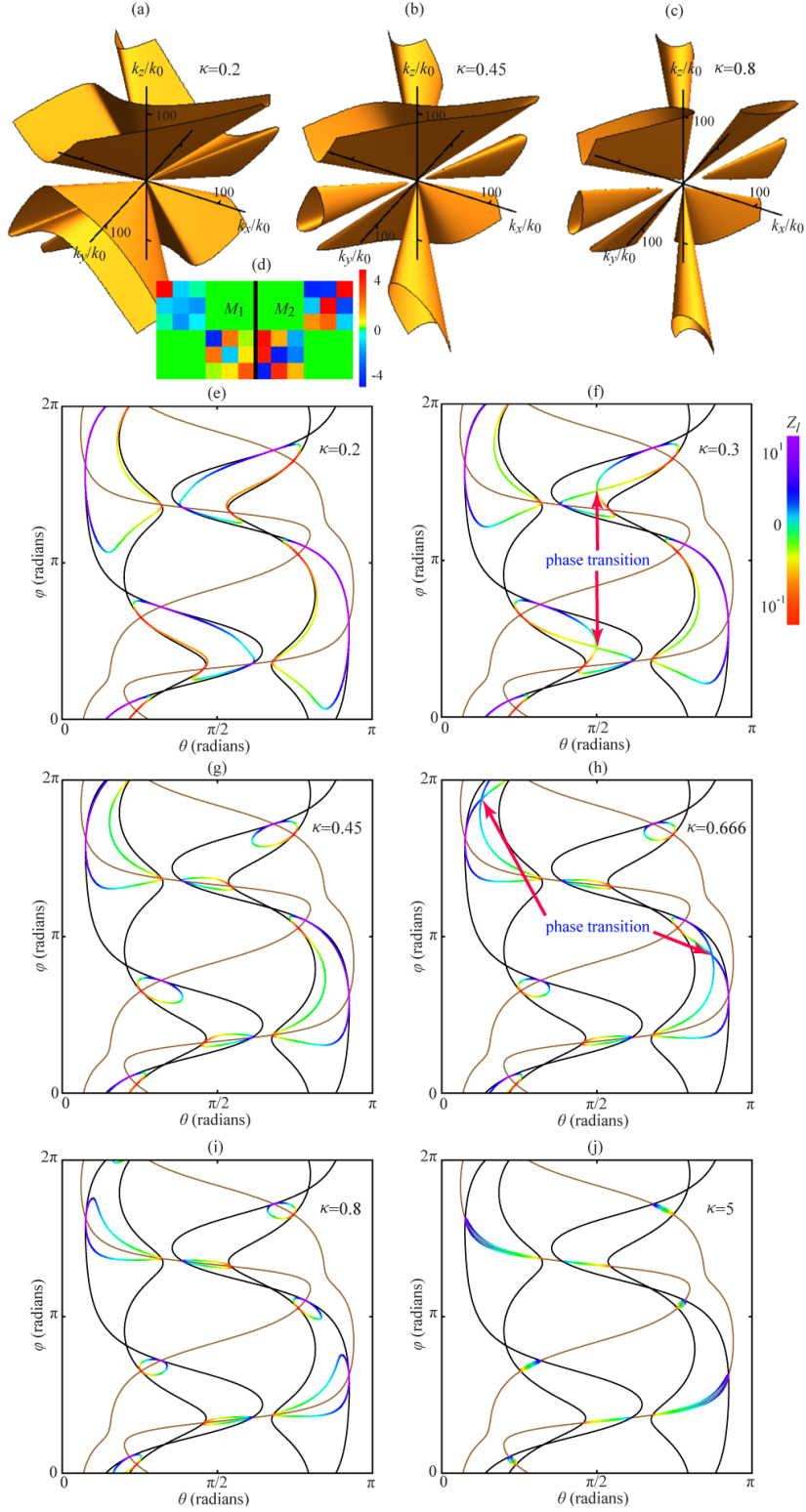

Fig. 3. Topological phase transitions (f),(h) between bi-hyperbolic [Fig. 3(a),(e)], tri-hyperbolic [Fig. 3(b),(g)] and tetra-hyperbolic [Fig. 3(c),(i)] phases with material parameters matrix $\widehat{M} = \widehat{M}_1 + i\kappa\widehat{M}_2$ as $\kappa$ is changed; matrices $M_1, M_2$ plotted on Fig. 3(d).

To conclude, let us ask the following question: what is the limit to the number of hyperboloids in iso-frequency surfaces of bi-anisotropic materials. To classify the quartic k-surfaces in the $k \gg k_0$ limit topologically, it is convenient to use the projective plane P2. As was noted in relation to bi-anisotropic materials [10] the classification follows from the following theorem [7]: A smooth projective real quartic curve consists topologically of: (i) one oval [hyperbolic medium]; (ii) two non-nested ovals [bi-hyperbolic medium]; (iii) two nested ovals; (iv) three ovals [tri-hyperbolic medium]; (v) four ovals [tetra-hyperbolic medium]; and (vi) an empty set (non-hyperbolic medium).

Therefore, in the present manuscript we complete the topological classification of bi-anisotropic optical materials.

**Funding.** Summer Research Session and College Office of Undergraduate Research (COUR) Grants from the College of Science and Mathematics at Georgia Southern University.